\documentclass{article}

\usepackage[final]{neurips_2020}

\usepackage[utf8]{inputenc} 
\usepackage[T1]{fontenc}    
\usepackage{hyperref}       
\usepackage{url}            
\usepackage{booktabs}       
\usepackage{amsfonts}       
\usepackage{nicefrac}       
\usepackage{microtype}      
\usepackage[utf8]{inputenc} 
\usepackage[T1]{fontenc}    
\usepackage{hyperref}       
\usepackage{url}            
\usepackage{booktabs}       
\usepackage{amsfonts}       
\usepackage{nicefrac}       
\usepackage{microtype}      
\usepackage{subfigure}
\usepackage{moreverb}
\usepackage{epsfig}
\usepackage{amsmath,amssymb,amsthm,mathrsfs,amsfonts,dsfont}
\usepackage{adjustbox,lipsum}
\usepackage{algorithm,algorithmic}
\usepackage{amsfonts}
\usepackage{epsfig}
\usepackage{amssymb}
\usepackage{amsmath}
\usepackage{xcolor}
\usepackage{algorithm,algorithmic}
\usepackage{multirow}
\usepackage{xcolor}     
\usepackage{mdframed}   
\usepackage{wrapfig} 
\usepackage{amsmath}
\usepackage{tcolorbox}
\usepackage{hyperref}

\makeatletter

\title{Generalized ADMM in Distributed Learning via Variational Inequality }

\author{%
	Saeedeh Parsaeefard   \,\,\,\,   \,\,\,\ \,\,\,\ Alberto Leon Garcia
	\\	Electrical and Computer Engineering\\
	University of Toronto\\
	Toronto, ON., M4P 1A6 \\
	\texttt{saeideh.fard,  alberto.leongarcia @utoronto.ca} \\
}

\begin{document}

\maketitle

\begin{abstract}
Due to the explosion in size and complexity of modern data sets and privacy concerns of data holders, it is increasingly important to be able to solve machine learning problems in distributed manners. The Alternating Direction Method of Multipliers (ADMM) through the concept of consensus variables is a practical algorithm in this context where its diverse variations and its performance have been studied in different application areas. In this paper, we study the effect of the local data sets of users in the distributed learning of ADMM. Our aim is to deploy variational inequality (VI) to attain an unified view of ADMM variations. Through the simulation results, we demonstrate how more general definitions of consensus parameters and introducing the uncertain parameters in distribute approach can help to get the better results in learning processes.  
\end{abstract}

\section{Introduction and Motivation}
Applications of machine learning (ML) are proliferating in many areas, e.g., Internet of things, smart factories, and health. These applications are built over highly distributed data gathering nodes, and the data sets are highly complex and large scale. As a result, it is highly desirable that the collection and storage of the data and accompanying algorithms be distributed. Distributed approaches are also promote the privacy of end-users who do not wish to share their data sets with third parties. In applied optimization theory, there is a large body of work facilitating distributed learning, such as primal-dual decomposition approaches, block-coordinate descent method, and alternating direction method of multipliers (ADMM). The latter has a simple but powerful structure \citep{distributedbook,BoydADMM} where for a set of local data owners/users/nodes/machines, i.e,  $\mathcal{N}=\{1,\cdots,N\}$, the following optimization problem is introduced
\begin{eqnarray} \label{ADMM1}
\min_{(\textbf{w}_n)_{n \in \mathcal{N}}, \textbf{z}}  \sum_{n \in \mathcal{N}}V_n(\textbf{w}_n, \textbf{x}_n, \textbf{y}_n), \,\,\, \,\,\,
\text{Subject to}:\,\,\, \text{C}1: \textbf{w}_n=\textbf{z}, \,\,\, \forall n \in \mathcal{N}, 
\end{eqnarray}
 {in which $\textbf{x}_n$ and $\textbf{y}_n$ are a feature set and a label set of user $n$, $\textbf{w}_n$ is local model of user $n$,} and $\textbf{z}$ is a common/consensus/global variable. This problem can be directly solved by its own augmented Lagrangian approach or simply as a special case of the constraint optimization problem \cite{BoydADMM}. ADMM is highly relevant to other distributed approaches such as proximal point method, projection method, and Douglas-Rachford splitting \citep{palomar}. Since ADMM can be fit into diverse categories of ML applications, it has been studied from different perspectives, e.g.,   in \citep{BoydADMM,COLA,9131841,7552562,6484993}. 
	
C1 in \eqref{ADMM1} is a tight constraint which causes slow convergence behavior of ADMM algorithm, and consequently, leads to extra message overhead among server and users. To overcome this implementation problem, there exist two main research streams in this literature: 1) Developing quantization and hierarchical methods, e.g., \citep{QGADMMbennis,9120758};2) Introducing relaxed versions of the consensus constrain in ADMM, e.g.,  \citep{francca2016explicit,10.5555/3045118.3045156,6892987,articleboyd}. Following by the former research trend, in this paper, we aim to introduce a more general version of the consensus constraints in ADMM for distributed learning to speed up the convergence rate. We also take into account the effect of distributed data sets of users in the performance of ADMM which to the best of our knowledge, has not been addressed in this literature. We introduce a worst case representation of the objective function in \eqref{ADMM1} where the effect of truncated /fragmented/local data sets of users is modeled as an unknown function which its values fall into the bounded uncertainty region \citep{federatedsaeedeh}.

{Since there exists no closed form solution of our proposed generalized ADMM, we use variational inequality (VI) \citep{PangVI} and arrive at a unified view for studying the proposed problem. We show how the solutions of ADMM in variations of the consensus constraints are related and how the worst case robust representation of ADMM can help to justify the regularization parameters in traditional approach. VI is also applied in \citep{ADMM22} for deriving divergence condition of ADMM, in \citep{non-convexAdmm} for the non-convex formulation of ADMM, and in \citep{8990114} for the convergence of ADMM with Bergman-distance formulation of C1 which can generalized ADMM for proximal and quadratic constraints. In this work, we show that VI can present an unified framework for study ADMM in general form of C1 without a need to have a closed form of mapping for the solution. }

\section{ Problem Representation for Generalized ADMM }

{The intuition behind introducing generalized formulation of ADMM is to allow users to search for the solution in a more flexible manner and study the effect of local data set of users in a distributed setup. Consequently,} we introduce the generalized ADMM formulation as follows for all $n$ in $\mathcal{N}$: 

\begin{eqnarray} \label{wcADMMfederatedlearning2}
\lefteqn{\min_{(\textbf{w}_n)_{n \in \mathcal{N}}, \textbf{z} } \sum_{n \in  \mathcal{N}} \max_{\textbf{f}_n(\textbf{w}_{-n},\textbf{x}_{-n}, \textbf{y}_{-n})  \in \Re_n(\textbf{w}_{-n}) } \widetilde{V}_n(\textbf{w}_n, \textbf{x}_n, \textbf{y}_n, \textbf{f}_n(\textbf{w}_{-n},\textbf{x}_{-n}, \textbf{y}_{-n}) ), \,\,\,} \\ &&
\quad \quad \,\,\,\text{Subject to}: \widetilde{\text{C1}}:\,\,\,  g_n^2(\textbf{w}_n, \textbf{z})\leq 0, \,\,\,  g_n^1(\textbf{w}_n, \textbf{z})=  0 ,  \,\,\, \forall n \in \mathcal{N}, \nonumber 
\end{eqnarray}
 {where $\textbf{w}_{-n}=[\textbf{w}_1, \cdots, \textbf{w}_{n-1}, \textbf{w}_{n+1}, \cdots \textbf{w}_{N} ]$, $\textbf{x}_{-n}=[\textbf{x}_1, \cdots, \textbf{x}_{n-1}, \textbf{w}_{n+1}, \cdots \textbf{w}_{N} ]$, and $\textbf{y}_{-n}=[\textbf{y}_1, \cdots, \textbf{y}_{n-1}, \textbf{y}_{n+1}, \cdots \textbf{y}_{N} ]$, and } 
\begin{itemize}
	\item $g_n^1(\textbf{w}_n, \textbf{z})$ and $g_n^2(\textbf{w}_n, \textbf{z})$ capture a more general relationship between $\textbf{w}_n$ and $ \textbf{z}$. For instance, $\widetilde{\text{C1}}$ can include norm function $p$ as $g_n^2(\textbf{w}_n, \textbf{z}) \quad \rightarrow \quad  \|\textbf{w}_n-\textbf{z}\|_p^p \leq \varepsilon_n $ for all $n$ for $\varepsilon_n \geq 0$. This type of soft constraints can be considered as an over relaxation of  {the consensus constraints} of ADMM approach (Chapter 3 in \citep{BoydADMM}). Another example is a group ADMM in \citep{GADMMbennis} where the users are agree to fix their solutions with their own neighbors, i.e., $\textbf{w}_n=\textbf{w}_{m}$ if $m \in \mathcal{A}_n$ where $\mathcal{A}_n$ is the adjacent list of the node $n$. 
\item  {Due to the local data sets in distributed setup, the obtained weights in learning process misses the relationship between data sets of users \citep{federatedsaeedeh}. Since there is no statistics that shows this effect, we need to define a new variable to model this relation through the worst case robust optimization theory which includes three elements: \textbf{{I)}} A new function to show the uncertainty coming from local data sets of users: Consider $\textbf{f}_{n}(\textbf{w}_{-n}, \textbf{x}_{-n}, \textbf{y}_{-n})$ which is a function to model the uncertainty caused by local date sets of users. Consequently, the cost function is also changed to $\widetilde{V}_n(\textbf{w}_n, \textbf{x}_n, \textbf{y}_n, \textbf{f}_n(\textbf{w}_{-n},\textbf{x}_{-n}, \textbf{y}_{-n}) )$ which includes $\textbf{f}_n(\textbf{w}_{-n},\textbf{x}_{-n}, \textbf{y}_{-n})$ as a new variable; \textbf{II)} An uncertainty region i.e., $\Re_n (\textbf{w}_{-n})$ to shows how $\textbf{f}_{n}(\textbf{w}_{-n}, \textbf{x}_{-n}, \textbf{y}_{-n})$ behaves: This uncertainty function is mathematically represented as \citep{protectionfunction1}
$$ \forall \, \textbf{w}_{-n},\textbf{x}_{-n}, \textbf{y}_{-n}, \,\,\,\,\, \exists	\, \textbf{f}_n(\textbf{w}_{-n},\textbf{x}_{-n}, \textbf{y}_{-n}) \in 
\Re_n (\textbf{w}_{-n}), \,\, \forall n \in \mathcal{N},$$
and it means for any weights, feature sets, and label sets of all users, a value of $\textbf{f}_{n}(\textbf{w}_{-n}, \textbf{x}_{-n}, \textbf{y}_{-n})$ is limited in $\Re_n (\textbf{w}_{-n})$ which is a set-valued map. For our problem, we do not have any knowledge about the form of $\Re_n (\textbf{w}_{-n})$. In Section 3.1, we use linear function for $\textbf{f}_n(\textbf{w}_{-n},\textbf{x}_{-n}, \textbf{y}_{-n}) $ and bounded norm function for $\Re_n (\textbf{w}_{-n})$; \textbf{III)} A mapping to show how the optimization problem should tackle the uncertain parameter: Definitely, by increasing the value of  $\textbf{f}_{n}(\textbf{w}_{-n}, \textbf{x}_{-n}, \textbf{y}_{-n})$, loss function increases. Therefore, under the worst-case or maximum condition of $\textbf{f}_n(\textbf{w}_{-n},\textbf{x}_{-n}, \textbf{y}_{-n})$ in $\Re_n(\textbf{w}_{-n})$, the weights of the model should be derived. Based on this view, the inner optimization problem of \eqref{wcADMMfederatedlearning2} is derived as
$$ {\Phi_n}=\max_{\textbf{f}_n  \in \Re_n(\textbf{w}_{-n}) } \widetilde{V}_n(\textbf{w}_n, \textbf{x}_n, \textbf{y}_n, \textbf{f}_n(\textbf{w}_{-n},\textbf{x}_{-n}, \textbf{y}_{-n}) ),$$}
 {where $\Phi_n$ is a function of all weights of users and depends on the bound of mathematical representation of $\Re_n(\textbf{w}_{-n}) $. For the sake of simplicity, we omit all the variables from the definition of this function.} 
\end{itemize}
 {The difficulty of finding a solution for \eqref{wcADMMfederatedlearning2} is twofold: 1) the objective of $\widetilde{V}_n(\textbf{w}_n, \textbf{x}_n, \textbf{y}_n, \textbf{f}_n(\textbf{w}_{-n},\textbf{x}_{-n}, \textbf{y}_{-n}) )$ is intertwined an uncertain function (i.e., $\textbf{f}_n(\textbf{w}_{-n},\textbf{x}_{-n}, \textbf{y}_{-n})$) and there is not a well-defined tractable formulation for this objective function, 2) the solution of each user from \eqref{wcADMMfederatedlearning2} depends on a shared variable $\textbf{z}$ among all users. To overcome these two challenges, we apply VI for studying the solution of \eqref{wcADMMfederatedlearning2}. VI has been successfully applied in the context of distributed algorithm with shared constraints, e.g., in \citep{non-convexAdmm,ADMM22} and can handle an ambiguity in definition of cost function in \eqref{wcADMMfederatedlearning2} by considering its  centralized counterpart problem  \citep{saeedehbook}, i.e.,   
\begin{eqnarray} \label{ADMMforCentralizedCounterpart}
\min_{\textbf{w}, \textbf{z} } \sum_{n \in  \mathcal{N}} V_n(\textbf{w}, \textbf{x}_n, \textbf{y}_n), \,\,\, \forall n \in \mathcal{N}, \,\, \text{Subject to}:  \widetilde{\text{C1}}, 
\end{eqnarray}}
where $\textbf{w}$ incorporates $\textbf{w}_n$ for all $n$.

\section{ Solution of Generalized ADMM}

To study the solution of Generalized ADMM with VI, we first present some preliminary definitions. 

\begin{tcolorbox}
	\textbf{Definition of VI problem:} Given a subset $\mathcal{K}$ of the Euclidean $n$-dimensional space $\mathbb{R}^{n}$ and a mapping $\mathcal{F}: \mathcal{K} \rightarrow \mathbb{R}^{n}$, the VI problem, denoted by $VI(\mathcal{K}, \mathcal{F})$, is to find a vector $\textbf{x} \in \mathcal{K}$ such that \citep{PangVI}
	$$ {(\textbf{y} -\textbf{x} )^{T}\mathcal{F}(\textbf{x}) \geq 0 , \,\,\,\,\, \forall \textbf{y} \in \mathcal{K}.}$$ 
	Let the solution set of above equation be $\text{SOL}(\mathcal{K}, \mathcal{F})$. VI can  represent the solution of several standard constraint/non-constraint optimization problems, equilibrium point of game theory, and complementarity problems. For instance, if $\mathcal{K}$ is convex and the mapping $\mathcal{F}$ is the gradient of real-valued function $f(\textbf{x})$, then $VI(\mathcal{K},  \mathcal{F}=\bigtriangledown f(\textbf{x}))$ represent a necessary condition for optimality of the following optimization problem: find $\textbf{x}^* \in \mathcal{K}$ such that $f(\textbf{x}^*) \leq f(\textbf{x})$ for all $\textbf{x} \in \mathcal{K}$. In case that $\mathcal{K}$ includes inequality and equations, i.e., $\mathcal{K}= \big\{\textbf{h}(\textbf{x}) \leq  \textbf{0} , \,\,\, \textbf{g}(\textbf{x})=\textbf{0} \,\,\,\forall n \in \mathcal{N}\big\}, $  with $\textbf{h}: \mathbb{R}^{n} \rightarrow  \mathbb{R}^{l} $ and  $\textbf{g}: \mathbb{R}^{n} \rightarrow  \mathbb{R}^{m} $ being vector-values continuously differentiable functions. The following two statements are valid [\citep{PangVI} Proposition 1.3.4.]:
	\begin{itemize}
		\item Consider $\textbf{x} \in \text{SOL}(\mathcal{K}, \mathcal{F})$, under Abadie's constraint qualifications (Section 3.2 in \cite{PangVI}), there exist positive vectors $\boldsymbol{\lambda} \in \mathbb{R}^{m}$ and $\boldsymbol{\mu} \in \mathbb{R}^{l}$ such that 
		\begin{eqnarray} \label{KKTofVI}
		\begin{array}{c}\textbf{0}=\mathcal{F}(\textbf{x}) + \sum_{j=1}^l \lambda_j\bigtriangledown h_j(\textbf{x})+ \sum_{i=1}^m \mu_i \bigtriangledown g_i(\textbf{x}), \,\, \forall n \in \mathcal{N},  \\
		\textbf{0}= \textbf{g}(\textbf{x}), \,\,\,\,
		\textbf{0} \leq \boldsymbol{\lambda} \bot \textbf{h}(\textbf{x}) \geq \textbf{0},\end{array}
		\end{eqnarray}
		where $\boldsymbol{\lambda} \bot \textbf{h}(\textbf{x})$ shows the perpendicularity of $\boldsymbol{\lambda}\geq \textbf{0}$ and $\textbf{h}(\textbf{x})$ where $\lambda_j \times h_j(\textbf{x})=0$. 
		\item Conversely, if each function ${h}_j$ is convex and each function $g_i$ is linear, and if $(\textbf{x},\boldsymbol{\lambda}, \boldsymbol{\nu} )$ satisfies  \eqref{KKTofVI}, then $\textbf{x} \in \text{SOL}(\mathcal{K}, \mathcal{F})$.	
	\end{itemize}  
	
\end{tcolorbox}

We refer to \citep{palomar,5447064} for more examples of using VI. From these preliminaries, we study the solution of generalized ADMM based on VI with the help of \eqref{ADMMforCentralizedCounterpart} as follows. Consider $\textbf{w}=\{\textbf{w}_1, \cdots \textbf{w}_N, \textbf{z}\}$ and the solutions of \eqref{wcADMMfederatedlearning2} and \eqref{ADMMforCentralizedCounterpart} as $\widetilde{\textbf{w}}$ and $\textbf{w}$, respectively. Assume also following practical assumptions\footnote{ {For the ease of presentation, in the followings, we omit the arguments from $\textbf{f}_n(\textbf{w}_{-n},\textbf{x}_{-n}, \textbf{y}_{-n})$. However, when we will explain about the concept, we add the arguments.}}:
\begin{itemize}
	\item \textbf{A1}) $V_n(\textbf{w}_n, \textbf{x}_n, \textbf{y}_n, \textbf{f}_n)$ is smooth, $L$ Lipschitz continuous, and  differentiable function of $\textbf{w}_n$ with bounded gradients; 
	\item  \textbf{A2}) $V_n(\textbf{w}_n, \textbf{x}_n, \textbf{y}_n, \textbf{f}_n)$ is increasing, bounded differentiable with respect to $\textbf{f}_{n}$;
	\item \textbf{A3}) $\frac{\partial^2 V_n(\textbf{w}_n, \textbf{x}_n, \textbf{y}_n, \textbf{f}_n)}{\partial \textbf{w}_n \partial \textbf{f}_n}$ exist and are continuous functions; 
	\item \textbf{A4}) Weights of users and $\textbf{z}$ are bounded i.e., $\textbf{w}\in \textbf{W}_n \subseteq \mathbb{R}^{n}$; \item  \textbf{A5}) Higher order of derivatives of $V_n(\textbf{w}_n, \textbf{x}_n, \textbf{y}_n, \textbf{f}_n)$ are approximately zero.
\end{itemize}
The solution of generalized ADMM in \eqref{wcADMMfederatedlearning2} can be represented as a solution of $VI(\widetilde{\mathcal{Q}}, \widetilde{\mathcal{F}})$ where $\widetilde{\mathcal{F}}=(\widetilde{\mathcal{F}}_n)_{n \in \mathcal{N}}$ in which $\widetilde{\mathcal{F}}_n= (\frac{\partial \Phi(\textbf{w})}{\partial \textbf{w}_n})$ 
and $$
\widetilde{\mathcal{Q}}=\prod_{n \in \mathcal{N}} \widetilde{\textbf{W}}_n\cap \big\{g_n^2(\textbf{w}_n, \textbf{z}) \leq  0 , \,\,\, g_n^1(\textbf{w}_n, \textbf{z})=0 \,\,\,\forall n \in \mathcal{N}\big\},$$ 
where $ \widetilde{\textbf{W}}_n=\big(\textbf{W}_n \times \Re_n(\textbf{w}_{-n})\big)$. Also, the solution of \eqref{ADMMforCentralizedCounterpart} can be represented as a solution of $VI(\mathcal{Q}, \mathcal{F})$ where $\mathcal{F}(\textbf{w})=(\mathcal{F}_n(\textbf{w}))_{n \in \mathcal{N}}$,  $\mathcal{F}_n(\textbf{w})=\nabla_{\textbf{w}} V_n(\textbf{w})$ and $\nabla_{\textbf{w}} V_n(\textbf{w})$ denotes the column gradient vector of $ V_n(\textbf{w})$ with respect to
$\textbf{w}$; and
\begin{equation}\label{Q}
\mathcal{Q}=\prod_{n \in \mathcal{N}}\textbf{W}_n \cap \big\{g_n^2(\textbf{w}_n, \textbf{z}) \leq  0 , \,\,\, g_n^1(\textbf{w}_n, \textbf{z})=0 \,\,\,\forall n \in \mathcal{N}\big\}. 
\end{equation}

\subsection{How to tackle Uncertain Parameters in Solution of generalized ADMM} Since \(  \)$\textbf{f}_n(\textbf{w}_{-n},\textbf{x}_{-n}, \textbf{y}_{-n})$ and $\Re_{n}$ are  {unknown}, it is not easy to determine a closed form or tractable formulation for $\widetilde{V}_n(\textbf{w}_n, \textbf{x}_n, \textbf{y}_n, \textbf{f}_n(\textbf{w}_{-n},\textbf{x}_{-n}, \textbf{y}_{-n}) )$ and consequently, $\widetilde{\mathcal{F}}(\textbf{w})$. To overcome this issue, we can apply two main approaches: \textbf{I)}
Considering simplified assumptions for the uncertainty region as well as the definition of cost functions with uncertain parameters. For example, assume that $\textbf{f}_n(\textbf{w}_{-n},\textbf{x}_{-n}, \textbf{y}_{-n}) )= \sum_{m \in \mathcal{N}, m \neq n }\alpha_{nm}\textbf{w}_m + \beta_{nm}$ where $\alpha_{nm}$ and $\beta_{nm}$ are uncertain parameters related to the correlation between the fragmented data sets of users $n$ and user $m$ and 
\begin{equation}\label{linearuncertaityfunction}
\Re_{n}= \{ \textbf{f}_n= \frac{1}{2} (\boldsymbol{\alpha}_{-n}^{T}\textbf{w}_{-n}-\boldsymbol{\beta}_{-n}) \, | \, 
\|\boldsymbol{\alpha}_{-n}\|_p\leq \varsigma_n, + \|\boldsymbol{\beta}_{-n}\|_{p} \leq \delta_{n} \}, \quad n \in \mathcal{N},
\end{equation}
where $\boldsymbol{\alpha}_{-n}$ and $\boldsymbol{\beta}_{-n}$ are two vectors of all elements of $\alpha_{nm}$ and $\beta_{nm}$, respectively.  {If $\widetilde{V}_n(\textbf{w}_n, \textbf{x}_n, \textbf{y}_n, \textbf{f}_n(\textbf{w}_{-n},\textbf{x}_{-n}, \textbf{y}_{-n}) )$} can be decomposed for user $n$ parameters and   $\textbf{f}_n(\textbf{w}_{-n},\textbf{x}_{-n}, \textbf{y}_{-n})$, by the concept of protection functions, i.e., $\widehat{V}_n(\textbf{f}_{n})$, we can have $\widetilde{V}_n (\textbf{w}_n, \textbf{x}_n, \textbf{y}_n, \textbf{f}_n(\textbf{w}_{-n},\textbf{x}_{-n}, \textbf{y}_{-n}) )\approx V_n(\textbf{w}_n, \textbf{x}_n, \textbf{y}_n,  )+\widehat{V}_n(\textbf{f}_{n}(\textbf{w}_{-n},\textbf{x}_{-n}, \textbf{y}_{-n}) ) $ for linear functions\footnote{  {or, $\widetilde{V}_n (\textbf{w}_n, \textbf{x}_n, \textbf{y}_n, \textbf{f}_n )\approx {V}_n(\textbf{w}_n, \textbf{x}_n, \textbf{y}_n, )\times\widehat{V}_n(\textbf{f}_{n}(\textbf{w}_{-n},\textbf{x}_{-n}, \textbf{y}_{-n}) )$ for the log or exponential functions.}}. These assumptions leas to a more tractable and closed form of both $\widetilde{V}_n (\textbf{w}_n, \textbf{x}_n, \textbf{y}_n, \textbf{f}_n(\textbf{w}_{-n},\textbf{x}_{-n}, \textbf{y}_{-n}) )$ and $\widetilde{\mathcal{F}}$. \textbf{II)} From A1-A4, we have following Lemma for $\mathcal{F}$ and $\widetilde{\mathcal{F}}$ which helps us to study the uniqueness condition of $\textbf{w}$ and $\widetilde{\textbf{w}}$.  

\textbf{Lemma}: When $\mathcal{F}$ is strongly monotone, $\widetilde{\mathcal{F}}$ is strongly monotone for small value of $\varsigma_n$ and from \textbf{A}1-\textbf{A}5. \textbf{\textit{Proof: }}  See Appendix A in supplementary file
$\square$. 

From the above definitions, we study the existence and uniqueness solution of ADMM. 


\textbf{Theorem 1:} For any set of parameters and bounded uncertainty region from A1-A4, there always exists a solution for $VI(\widetilde{\mathcal{Q}}, \widetilde{\mathcal{F}})$ and $VI(\mathcal{Q}, \mathcal{F})$. Therefore, \eqref{wcADMMfederatedlearning2} and \eqref{ADMMforCentralizedCounterpart} have a solution.  
\textit{\textbf{Proof:}} See Appendix B in supplementary file $\square$

While the existence condition is studied in a straightforward manner from the definitions of VIs related to  \eqref{wcADMMfederatedlearning2} and \eqref{ADMMforCentralizedCounterpart}, the uniqueness condition is not trivial since the uniqueness conditions depend on the nature of constraints. To consider this point, we use KKT systems of VI for deriving the uniqueness conditions. Consider $\textbf{g}^2$ and $\textbf{g}^1$ as the vectors of all $g_n^2$ and $g_n^1$, respectively.The KKT condition system of equations of \eqref{wcADMMfederatedlearning2} and \eqref{ADMMforCentralizedCounterpart} are, respectively,

\begin{eqnarray} \label{LagrnageADMMforCentralizedCounterpart2}
\begin{array}{c}\textbf{0}=\widetilde{\mathcal{F}}(\textbf{w}) + \sum_{n \in \mathcal{N}} (\mu_n\bigtriangledown g_n^1(\textbf{w}_n, \textbf{z})+ \lambda_n \bigtriangledown g_n^2(\textbf{w}_n, \textbf{z})), \,\, \forall n \in \mathcal{N},  \\
\textbf{C2}: \quad \textbf{0}= \textbf{g}^1(\textbf{w}_n, \textbf{z}), \,\,\,\,
\textbf{0} \leq \boldsymbol{\lambda} 	\bot \textbf{g}^2(\textbf{w}_n, \textbf{z})) \geq \textbf{0},\end{array}
\end{eqnarray}
and 
\begin{eqnarray} \label{LagrnageADMMforCentralizedCounterpart}
\begin{array}{c}\textbf{0}=\mathcal{F}(\textbf{w}) + \sum_{n \in \mathcal{N}} \mu_n\bigtriangledown g_n^1(\textbf{w}_n, \textbf{z})+ \lambda_n \bigtriangledown g_n^2(\textbf{w}_n, \textbf{z}) \,\, \forall n \in \mathcal{N},  \quad \textbf{C2}. \end{array}
\end{eqnarray}
\eqref{LagrnageADMMforCentralizedCounterpart2} and \eqref{LagrnageADMMforCentralizedCounterpart} belong to MiCP and provide a set of necessary conditions which should be satisfied for any solutions of  $VI(\widetilde{\mathcal{Q}}, \widetilde{\mathcal{F}})$ and $VI(\mathcal{Q}, \mathcal{F})$, respectively. Consider $(\boldsymbol{\lambda}, \boldsymbol{\mu})$ satisfying \eqref{LagrnageADMMforCentralizedCounterpart2} and \eqref{LagrnageADMMforCentralizedCounterpart} ; and $\widetilde{\textbf{w}}$ and $\textbf{w}$ are primal variables KKT points of \eqref{LagrnageADMMforCentralizedCounterpart2} and \eqref{LagrnageADMMforCentralizedCounterpart} , respectively. Now from Proposition 1.2.1 in \citep{PangVI}, $\textbf{w}$  is a solution of \eqref{ADMMforCentralizedCounterpart} from $VI(\mathcal{Q}, \mathcal{F})$ if there exist $\boldsymbol{\mu}$ and $\boldsymbol{\lambda}$ holding \eqref{LagrnageADMMforCentralizedCounterpart}. Similarly, $\widetilde{\textbf{w}}$ is a solution of \eqref{wcADMMfederatedlearning2} from $VI(\widetilde{\mathcal{Q}}, \widetilde{\mathcal{F}})$ if there exist $\boldsymbol{\mu}$ and $\boldsymbol{\lambda}$ holding $\textbf{C}2$. This relation helps us to study the uniqueness condition of general ADMM where 
associated with $VI(\mathcal{Q}, \mathcal{F})$ and  $VI(\widetilde{\mathcal{Q}}, \widetilde{\mathcal{F}})$. The solution of  \eqref{LagrnageADMMforCentralizedCounterpart2} and \eqref{LagrnageADMMforCentralizedCounterpart} can be represented as tuples $\widetilde{\boldsymbol{\varkappa}}=[\widetilde{\textbf{w}},\boldsymbol{\lambda}, \boldsymbol{\mu}]$ and  $\boldsymbol{\varkappa}=[\textbf{w},\boldsymbol{\lambda}, \boldsymbol{\mu}]$, respectively. Now we define two following mappings
\begin{eqnarray}\label{Lagrange1AVI}
\widetilde{\mathcal{F}}^{\text{Lagrange}}(\widetilde{\boldsymbol{\varkappa}})=\left( \begin{array}{c}\widetilde{\mathcal{F}}(\textbf{w})+\mathcal{G}^1 \boldsymbol{\mu}^{T}+\mathcal{G}^2 \boldsymbol{\lambda}^{T} \\\textbf{g}^1\\\textbf{g}^2
\end{array} \right), \,\, \text{and} \,\, \mathcal{F}^{\text{Lagrange}}(\boldsymbol{\varkappa})=\left( \begin{array}{c}\mathcal{F}(\textbf{w})+\mathcal{G}^1 \boldsymbol{\mu}^{T}+\mathcal{G}^2 \boldsymbol{\lambda}^{T} \\\textbf{g}^1\\\textbf{g}^2
\end{array} \right) \,\,\nonumber 
\end{eqnarray}
where $\mathcal{G}^1$ and $\mathcal{G}^2$ are the vectors of $\bigtriangledown g_n^1(\textbf{w}_n, \textbf{w})$ and $\bigtriangledown g_n^2(\textbf{w}_n, \textbf{w})$, respectively. $\widetilde{\mathcal{F}}^{\text{Lagrange}}(\widetilde{\boldsymbol{\varkappa}})$ and $\mathcal{F}^{\text{Lagrange}}(\boldsymbol{\varkappa})$  can be used to define a new mapping between the primal and Lagrange variables which can be used to study the uniqueness conditions of  $\widetilde{\textbf{w}}$ and $\textbf{w}$.

\subsection{Simplified Version for Linear Constraints}

Another important advantage to use VI is that it is simplified when  $\textbf{g}^1=\textbf{d}-\textbf{C}\textbf{w}$ and  $\textbf{g}^2=\textbf{b}-\textbf{A}\textbf{w}$. In this case,  $\mathcal{F}^{\text{Lagrange}}$ and $\widetilde{\mathcal{F}}^{\text{Lagrange}}$ are transformed into
\begin{eqnarray}\label{15}
\mathcal{F}^{\text{Linear}}(\boldsymbol{\varkappa})=\underbrace{\left( \begin{array}{c}\textbf{0} \\\textbf{d}\\\textbf{b}
	\end{array} \right)}_{\textbf{m}}+\underbrace{\left( \begin{array}{ccc}\mathcal{F} & \textbf{C}^T & \textbf{A}^T \\-\textbf{C}^T & \textbf{0} & \textbf{0}\\-\textbf{A}^T & \textbf{0} & \textbf{0}
	\end{array} \right)}_{\textbf{M}}\boldsymbol{\varkappa}^T,  \,  \widetilde{\mathcal{F}}^{\text{Linear}}(\widetilde{\boldsymbol{\varkappa}})=\textbf{m}+\underbrace{\left( \begin{array}{ccc}\widetilde{\mathcal{F}} & \textbf{C}^T & \textbf{A}^T \\-\textbf{C}^T & \textbf{0} & \textbf{0}\\-\textbf{A}^T & \textbf{0} & \textbf{0}
	\end{array} \right)}_{\widetilde{\textbf{M}}}\boldsymbol{\varkappa}^T, \nonumber 
\end{eqnarray}
where $\textbf{M}$ and $\widetilde{\textbf{M}}$ can be rewritten as $\textbf{M}=\textbf{M}_1+\textbf{M}_2$ and $\widetilde{\textbf{M}}=\widetilde{\textbf{M}}_1+\textbf{M}_2$ where $$\textbf{M}_1=\left( \begin{array}{ccc}\mathcal{F} & \textbf{0} & \textbf{0} \\\textbf{0} & \textbf{0} & \textbf{0}\\\textbf{0} & \textbf{0} & \textbf{0}
\end{array} \right), \,\,\,\widetilde{\textbf{M}}_1=\left( \begin{array}{ccc}\widetilde{\mathcal{F}} & \textbf{0} & \textbf{0} \\\textbf{0} & \textbf{0} & \textbf{0}\\\textbf{0} & \textbf{0} & \textbf{0}
\end{array} \right),\,\,\,\,\,\, \textbf{M}_2=\left( \begin{array}{ccc} \textbf{0}& \textbf{C}^T & \textbf{A}^T \\-\textbf{C}^T & \textbf{0} & \textbf{0}\\-\textbf{A}^T & \textbf{0} & \textbf{0}
\end{array} \right)$$
in which $\textbf{M}_2$ is a skew-symmetric matrix. For instance, for $\textbf{w}_n=\textbf{z}$, we have $\textbf{z}=-\frac{1}{N}\sum_{n \in \mathcal{N}}\textbf{w}_n$ \citep{BoydADMM}. Hence, $\textbf{C}$ is an $N \times N$ matrix $[C]_{nm}= 1-\frac{1}{N} $ for $m=n$ and $[C]_{nm}= \frac{1}{N} $ for $m\neq n$. For group ADMM in \citep{GADMMbennis}, $\textbf{C}$ is $N \times N$ is matrix where $[C]_{nm}=1$ for $m=n$, $[C]_{nm}=-1$ for $m \neq n$ and $n \in \mathcal{A}_n $, and $[C]_{nm}=0$, otherwise. For both of these two cases, the study of the existence, uniqueness, and convergence of distributed algorithms depends on the features of $\mathcal{F}$ in ADMM and not the linear matrices.

 {Applying VI in ADMM also helps us to study any variation of ADMM including adding any additional terms to the cost functions, for example (to punish or optimize the relaxation parameters, e.g., $\varepsilon_n$}) or any new constraints and variables, and make a connection between solution of simplified form and its divergence. Therefore, we get a unified view for study the solution of ADMM for different scenarios. 

\section{Uniqueness Conditions of GADMM and Distributed Algorithms} 
We first focus on the case that $\boldsymbol{\lambda}$ and $\boldsymbol{\mu}$ are fixed. This is interesting in a sense that it leads to a practical formulation of augmented Lagrange formulation in ADMM scenario and can be considered as a first step for the scenario where the values of Lagrange multipliers are adjusted dynamically. 


\subsection{Fixed Lagrange Multipliers }

For this case, the constraints of the KKT systems in (7) and (8) are eliminated and we have 
$\mathcal{F}^{\text{Fixed}}(\textbf{w})= \mathcal{F}(\textbf{w})+\sum_{n \in \mathcal{N}} \left(\mu_n\bigtriangledown g_n^1(\textbf{w}_n, \textbf{z})+ \lambda_n \bigtriangledown g_n^2(\textbf{w}_n, \textbf{z})\right)
$,
and 
$\widetilde{\mathcal{F}}^{\text{Fixed}}(\textbf{w})= \widetilde{\mathcal{F}}(\textbf{w})+\sum_{n \in \mathcal{N}} \left(\mu_n\bigtriangledown g_n^1(\textbf{w}_n, \textbf{z})+ \lambda_n \bigtriangledown g_n^2(\textbf{w}_n, \textbf{z})\right)$,
which can represent two following optimization problems :$$\min_{(\textbf{w}_n)_{n \in \mathcal{N}} \in \textbf{W}} \sum_{n \in  \mathcal{N}} \left(V_n(\textbf{w}, \textbf{x}_n, \textbf{y}_n)+\mu_n g_n^1(\textbf{w}_n, \textbf{z})+\lambda_ng_n^2(\textbf{w}_n, \textbf{z})\right), $$ 
$${\min_{(\textbf{w}_n)_{n \in \mathcal{N}} \in \textbf{W}} \sum_{n \in  \mathcal{N}} \left(\max_{\widehat{\textbf{f}}_n  \in \Re_n(\textbf{w}_{-n}) } \widetilde{V}_n(\textbf{w}_n, \textbf{x}_n, \textbf{y}_n, \textbf{f}_n(\textbf{w}_{-n},\textbf{x}_{-n}, \textbf{y}_{-n}) ) +\mu_n g_n^1(\textbf{w}_n, \textbf{z})+\lambda_ng_n^2(\textbf{w}_n, \textbf{z})\right)},$$respectively. These two optimization problems show that \eqref{wcADMMfederatedlearning2}and \eqref{ADMMforCentralizedCounterpart} are transformed into  optimization problems with the objective to minimize the cost function plus the pricing parts (fixed Lagrange multipliers multiplied by $\textbf{g}_1$ and $\textbf{g}_2$). Therefore, the related optimization problems are simplified and their related VI mappings are changed to $VI(\textbf{W}, \mathcal{F}^{\text{Fixed}})$ and  $VI(\widetilde{\textbf{W}}, \widetilde{\mathcal{F}}^{\text{Fixed}})$. 

\textbf{Theorem 2}:  For a given $\boldsymbol{\lambda} \geq \textbf{0}$ and $\boldsymbol{\mu} \geq \textbf{0}$, when $\varsigma_n$ is small and $g_n^1$ is linear and $g_n^2$ is convex function, $VI(\textbf{W}, \mathcal{F}^{\text{Fixed}})$ and  $VI(\widetilde{\textbf{W}},\widetilde{ \mathcal{F}}^{\text{Fixed}})$ have  unique solutions if $\mathcal{F}(\textbf{w})$ is strongly monotone. \textbf{\textit{Proof: }}  See Appendix C in supplementary file
$\square$.

For this scenario, the iterative prototype algorithm based on the proximal response point can be applied as presented in Algorithm 1 \citep{Palomarbook}. In proximal point method, for any $\textbf{w}_n \in \mathcal{Q}$, the solution of   
\begin{equation}\label{proxi12}
\textbf{w}_n^{t}= \min_{\textbf{w}_n}\widetilde{V}_n(\textbf{w}_n, \textbf{x}_n, \textbf{y}_n, \textbf{f}_n ) +\mu_n\bigtriangledown g_n^1(\textbf{w}_n, \textbf{z})+ \lambda_n \bigtriangledown g_n^2(\textbf{w}_n, \textbf{z}) +\underbrace{\frac{1}{2}\|\textbf{w}^t_n-\textbf{w}^{t-1}_n\|}_{\text{Proximal term}}.
\end{equation}
For $\textbf{z}$, the similar proximal term can be considered. This algorithm is based on the sequential distributed algorithm where all users starts to solve their problems by sending their solution back to the centralized cloud. With a very slight modification, this algorithm can be applied for the server-less algorithm. 

\begin{table*}[]
	\caption{Proximal Point Method for GADMM with Fixed  and Variable Lagrange Multipliers } \centering \vspace{-0.0 in}
\begin{tabular}{ll}
		\begin{tabular}{l}
		\hline \hline 
		\textbf{Algorithm 1}
		\\
		\textbf{Step 0}: For given $\boldsymbol{\lambda}$, $\boldsymbol{\mu}$, and $\textbf{z}^0$, $0<\zeta<< 1$, $t=0$
		\\  \textbf{Step 1}: 1.1) Users update $\textbf{w}_n^{t+1}$  from \eqref{proxi12} 	
		\\ \quad \quad \quad $\,$  1.2) Each user sends its $\textbf{w}_n^{t+1}$ to the server 		
		\\ \textbf{Step 2}: The server updates 	
		$ \textbf{z}^{t+1}$ and sends to users 
		\\
		$\quad \quad \quad $If $\|\textbf{w}^{t-1}-\textbf{w}^{t}\|_{2}\leq \zeta$, End;\\ Otherwise $t=t+1$, continue;
		\\ \hline \hline
		\vspace{-0.0 in}
	\end{tabular}
	&
		\begin{tabular}{l}
		\hline \hline 
		\textbf{Algorithm 2}\\
				\textbf{Step 0}:  {For given $0<\zeta, \tau << 1$, $\textbf{z}^0$, $t=0$,}
		\\  \textbf{Step 1.1 }) similar to Algorithm 1 
		\\  \textbf{Step 1.2}) similar to Algorithm 1 	
		\\ 
		1.3) Update $\boldsymbol{\lambda}^t$ and $\boldsymbol{\mu}^t$
		\\ $\,\,\,$ from one of \eqref{Projection} to  \eqref{Tikhonoc}
			\\ \textbf{Step 2}: Similar to Algorithm 1
		\\ \hline \hline
		\vspace{-0.0 in}
	\end{tabular}
\end{tabular}
\end{table*}
Consider the norm 2 uncertainty region in \eqref{linearuncertaityfunction} with the linear regression cost function and $\textbf{C}1$ for ADMM from \eqref{ADMM1}, the protection function against $\textbf{f}_n$ is $\varsigma_n \|\textbf{w}_{-n}^t\|_2^2+\delta_n$ for each user $n$, and the optimization problem of proximal response map for each user $n$ can be rewritten 
\begin{eqnarray}\label{proxilinear}
V_n^{\text{Proxi}}=\underbrace{\frac{1}{2} (\textbf{x}_n^{T}\textbf{w}_n^{t}-\textbf{y}_n)}_{V_n(\textbf{w}_n,\textbf{x}_n,\textbf{y}_n)}+\underbrace{\mu_n (\textbf{w}_n^{t}- \textbf{z})}_{\text{ADMM Constraint}}+\underbrace{\varsigma_n \|\textbf{w}_{-n}^t\|_2^2+\delta_n+ \frac{1}{2}  \|\textbf{w}_n^t-\textbf{w}_{n}^{t-1}\|_2^2}_{\text{Regularization Paramter}}. 
\end{eqnarray}
\eqref{proxilinear} is an augmented version of ADMM via the weights of each user as well as the other users and $\varsigma_n$ and $\delta_n$. \eqref{proxilinear} highlights that with representation of distributed learning via worst case optimization theory based on ADMM and proximal response method, the utilization of regularization factor in ADMM approach can be supported. With different definition of uncertainty region, we obtain a different regularization factor. Hence, this representation mathematically supports applying regularization factors in ADMM to reach the better solutions. Convergence conditions of Algorithm 1 are presented in Appendix D in supplementary file.





\subsection{Variable Lagrange Multipliers }
For this scenario, we just focus on $\boldsymbol{\lambda}$ for the sake of notation simplicity. The followings can be easily extended with considering $\boldsymbol{\mu}$ as well. Therefore, we study that under which condition for the unique solution of $\textbf{w}$, there is a solution for $\boldsymbol{\lambda}$ which hold KKT systems for both \eqref{ADMMforCentralizedCounterpart} and  \eqref{wcADMMfederatedlearning2}, and  $\boldsymbol{\lambda}$ is a solution of 
\begin{eqnarray}\label{NCP}
\text{NCP}:    \quad \textbf{0} \leq \boldsymbol{\lambda} \perp \textbf{g}^2(\bar{\boldsymbol{\varkappa}}) \geq \textbf{0},
\end{eqnarray}
where $\bar{\boldsymbol{\varkappa}}=\{\textbf{w}, \boldsymbol{\mu}\}$. 
To study this uniqueness conditions for $\bar{\boldsymbol{\varkappa}}$ and $\boldsymbol{\lambda}$, consider $\boldsymbol{\gamma}(\boldsymbol{\lambda})=\mathcal{G}^2 \boldsymbol{\lambda}^{T} $ and define a mapping $\boldsymbol{\phi}$ as $
\boldsymbol{\phi}: \,\,  \boldsymbol{\lambda} \rightarrow \textbf{g}^2(\bar{\boldsymbol{\varkappa}},\gamma(\boldsymbol{\lambda}))$. For this scenario, we need to show that under which condition $\boldsymbol{\phi}$ is co-coercive function of $\boldsymbol{\lambda}$ with constant $c_{\text{coc}}(\boldsymbol{\phi})$ where   $\big(\boldsymbol{\lambda}^1-\boldsymbol{\lambda}^2)^T(\boldsymbol{\phi}(\boldsymbol{\lambda}^1)-\boldsymbol{\phi}(\boldsymbol{\lambda}^2))\geq c_{\text{coc}}(\boldsymbol{\phi})\|\boldsymbol{\phi}(\boldsymbol{\lambda}^1)-\boldsymbol{\phi}(\boldsymbol{\lambda}^2)\|_2^2 $.

\textbf{Theorem 3}: If $\mathcal{F}$ is strong monotone, the solution of $VI(\mathcal{Q}, \mathcal{F})$ and  $VI(\widetilde{\mathcal{Q}}, \widetilde{\mathcal{F}})$ are unique and it must be $\textbf{w}^{VI}=\textbf{w}^*(\boldsymbol{\gamma}(\boldsymbol{\lambda}^{\text{NCP}}))$ and $\widetilde{\textbf{w}}^{VI}=\widetilde{\textbf{w}}^*(\boldsymbol{\gamma}(\boldsymbol{\lambda}^{\text{NCP}}))$. There is a convex solution set for $\boldsymbol{\lambda}$. \textit{\textbf{Proof:}} See Appendix E in supplementary file. 
 $\square$

\subsubsection{Distributed Algorithms to Solve \eqref{NCP}}
One scheme of a distributed algorithm for \eqref{NCP} is presented in Algorithm 2 which is based on the primal-dual update variables.  {First, server  each user updates its own local variable $\textbf{w}_n$ based on initial values of $\textbf{z}^0$ . Then, each user should update the Lagrange multipliers which involves different implementation challenges such as: 1) The signaling required between server and users to pass required information, 2)	The computational burden for calculating $\boldsymbol{\lambda}$, 3) The convergence speed,  4) The convergence analysis, 5) The sensitivity of the solution for $\boldsymbol{\lambda}$ to the step-size. Based on the definition of consensus general constraints in our setup, e.g.,  ${g}_n^1(\textbf{w}_n, \textbf{z})$ and ${g}_n^2(\textbf{w}_n, \textbf{z})$, each user can update $\lambda_n^t$ and $\mu_n^t$. However, computational complexity and choosing the step size for updating these parameters are important for $\textbf{w}^t$. We study some of the famous approaches to update the Lagrange multiplier as follows \citep{PangVI,5411806}.}

\begin{itemize}
	\item \textbf{ \textit{Projection Algorithm with random variable step}} where $\boldsymbol{\lambda}$ are updated as follows
	\begin{eqnarray}\label{Projection}
	\boldsymbol{\lambda}^{t}=[\boldsymbol{\lambda}^{t-1}- \tau	\boldsymbol{\phi}( \boldsymbol{\lambda}) ]^+,
	\end{eqnarray}
	where  {to guarantee the convergence of the algorithm, $0 \leq \tau \leq 2 \times c_{\text{coc}}(\boldsymbol{\phi})$ (Theorem 12.1.8 in \citep{PangVI})). Since this value usually cannot be attained in a distributed setup, very small value for $\tau$ is considered with the cost of slow convergence rate. The positive side of this approach is that it does not involve high computational complexity.} 
	\item 
	\textbf{\textit{The Hyperplane Projection Algorithm}}  is proposed to handle the difficulty of choosing the value of $\tau$ in the projection mapping. This algorithm has three steps to update the Lagrange multipliers: 
\begin{enumerate}
	\item Compute $
	\boldsymbol{\lambda}^{(t-1)+\frac{1}{4}}=[\boldsymbol{\lambda}^{t-1}- \boldsymbol{\phi}( \boldsymbol{\lambda}) ]^+$.\item Compute $\boldsymbol{\lambda}^{t-\frac{1}{2}}=\boldsymbol{\lambda}^{t-1}+2^{-l_n}(\boldsymbol{\lambda}^{(t-1)+\frac{1}{4}}-\boldsymbol{\lambda}^{t-1})$ where $l_n$ is the smallest positive integer $l$ where $\tau=2^{-l}$, and for $\delta\in (0,1)$, we have $[\boldsymbol{\lambda}^{t-1}-\boldsymbol{\lambda}^{(t-1)+\frac{1}{4}}] \boldsymbol{\phi}(\boldsymbol{\lambda}^{t-1}-\tau (\boldsymbol{\lambda}^{t-1}-\boldsymbol{\lambda}^{(t-1)+\frac{1}{4}}) \geq \delta \|\ \boldsymbol{\lambda}^{t-1}-\boldsymbol{\lambda}^{(t-1)+\frac{1}{4}}\|$.
	\item Update \begin{equation}
	\boldsymbol{\lambda}^{t}= \big[ \boldsymbol{\lambda}^{t-1} - \frac{[\boldsymbol{\lambda}^{t-1}-\boldsymbol{\lambda}^{(t-1)+\frac{1}{4}}]\boldsymbol{\phi}(\boldsymbol{\lambda}^{t-\frac{1}{2}})}{\|\boldsymbol{\phi}(\boldsymbol{\lambda}^{t-\frac{1}{2}})\|^2}\boldsymbol{\phi}(\boldsymbol{\lambda}^{t-\frac{1}{2}}) \big]^+.
	\end{equation}
\end{enumerate}
 {Obviously this approach has more computational complexity compared to the projection algorithm and there is no guarantee for the uniqueness solution for $\boldsymbol{\lambda}$.} 
\item \textbf{\textit{The Tikhonoc Regularization Algorithm}} is which instead of $\boldsymbol{\phi}$, its perturbed version, i.e.,  $\boldsymbol{\phi}+\zeta_n\textbf{I}$ is solved where $\textbf{I}$ is the identity matrix and $\zeta_n$ is a small value which is decreased during the sequence. Therefore, the updating of the Lagrange multipliers for each $t$ is changed as follows: 
\begin{eqnarray}\label{Tikhonoc}
\widehat{\boldsymbol{\lambda}}_n^{l+1,t}=\big[\widehat{\boldsymbol{\lambda}}^{l,t}- \frac{1}{\tau_n}(\boldsymbol{\phi}( \boldsymbol{\lambda}^t) +\zeta_n\widehat{\boldsymbol{\lambda}}^{l,t})\big]^+
\end{eqnarray} 
 {where  $l=0, 1, \cdots$ and this algorithm converges when $\tau_n > \frac{(\frac{1}{c_{\text{coc}}}+\zeta_n)^2}{2\zeta_n}$. This approach has an advantage to attain the unique least-norm solution of the NCP($\boldsymbol{\phi}$) for each $t$. } 
\end{itemize}
Algorithm 2 continuous until the convergence condition holds. Under strong monotonicty of $\mathcal{F}$, their convergence  are studied in Theorems 12.1.8, 12.1.16, and 12.1.2 in \citep{PangVI}, respectively.  

\section{Evaluation Results}
To evaluate the generalized ADMM, we deploy the setup of Matlab codes from \citep{BoydADMM,GADMMbennis}. We evaluate the performance in terms of numbers of iterations for convergence and  {we define $\Delta$ which is the difference between solution in \eqref{wcADMMfederatedlearning2} and \eqref{ADMMforCentralizedCounterpart}.} The distributed algorithm will be stopped when the difference between the solution of consecutive iterations is less than $10^{-4}$. We simulate for both logistic and Linear regression objective over the data set 29 in \citep{GADMMbennis}. We use the projection algorithm to update the value of $\boldsymbol{\lambda}$ where we have $\tau=0.0002$. For the protection function, we use norm 2, i.e., $\varsigma_n \|\textbf{w}_{-n}^t\|_2^2+\delta_n$ where we set  $\varsigma_n=5 \times 10^{-3}$ and $\delta_n=0.1$. 

The effect of soft constraints in the distributed solution of ADMM is studied in Table \ref{tab:multicol}. The results reveal that by increasing the value of $\varepsilon=\varepsilon_n$ for all $n \in \mathcal{N}$, the error is increased since the users are allowed to choose the solution close to each other and not exactly similar to each other. However, interestingly, the convergence time is decreased significantly since users can choose the solution that is in proximity of the consensus variables. Therefore, soft constraints allow us to make a trade-off between convergence and error.

Note that we examine the solution of ADMM for $\varepsilon<0.02$ and $\varepsilon<0.5$. In the former scenario, $\Delta$ is approaching zero and there is no significant difference between soft constraints and the  traditional consensus constraint. For the latter scenario, the error is not acceptable and the solution is not within an acceptable bound of accuracy. For linear regression, the error value is normalized with the value of the error of gradient descent method. The trend of increasing $\Delta$ and decreasing iteration numbers with increasing $\varepsilon$ are similar to those of logistic regression.

For large value of $\varepsilon$, i.e., $\varepsilon=0.5$, the error of ADMM with soft constraint is larger than that of the gradient algorithm, while it has a very fast convergence. Note that to reach the convergence for large value of $\varepsilon$, we decrease the steps sizes of the iterative parameters. Since in distributed setting for the large value of $\varepsilon$, due to flexibility of the choosing the values of weights of users, there is a chance that the weights of users oscillate. Therefore, there is a need to adjust the learning parameters in this scenarios with small values for better convergence.  


\begin{table}[ht]
	\begin{center}
		\caption{Performance of ADMM in presence of soft constraint versus $\varepsilon=\varepsilon_n$ for all $n \in \mathcal{N}$   }
		\begin{tabular}{cccccccc}
			\hline
			&	\multicolumn{3}{c}{Logistic Regression }&\multicolumn{3}{c}{Linear Regression}\\
			
			&	$\varepsilon=0.05$&$\varepsilon=0.1$&$\varepsilon=0.5$&$\varepsilon=0.01$&$\varepsilon=0.05$&$\varepsilon=0.5$ \\ \hline \hline \\
			$\Delta$&	3.8 $\times 10^{-3}$&1.5 $\times 10^{-2}$&0.03&0.80\%&0.95\%& 135\%\\
			Iterations &6000&3000&1700&300&200& 16\\
			
			\hline
		\end{tabular}
		\label{tab:multicol}
	\end{center}
\end{table}

\begin{figure*}
	\centering
	\includegraphics[width=3 in]{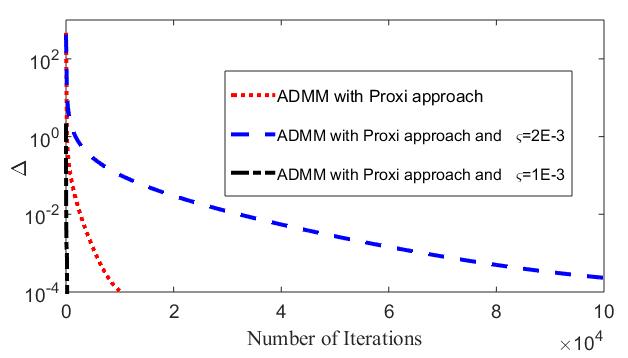}
	\caption{$\Delta$ versus number of iterations for  logistic regression and L2 as a protection function.}
	\label{ForpaperADMM}
\end{figure*}

In Fig. \ref{ForpaperADMM}, we study the effect of the protection function and the value of $\varsigma$ on the performance of proximal point method for generalized ADMM (proxi-ADMM) in Algorithm 2 for logistic regression problem. We plot the values of $\Delta$ versus iteration numbers for $\varsigma=10^{-3}$ and $\varsigma=2\times 10^{-3}$. The protection function is $L2$ norm. Fig. \ref{ForpaperADMM} shows that $\varsigma=10^{-3}$ improves the performance of the proxi-ADMM in terms of $\Delta $ and convergence rate compared to those for proxi-ADMM without the protection function. However, for $\varsigma=2\times 10^{-3}$, we cannot see the same improvement. This results indicates the importance of adjusting the value of protection functions and its non-uniform behavior in terms of improving performance as reported in similar trend of research e.g., \citep{federatedsaeedeh}.

\section{Conclusion}
 {We study the generalized ADMM for distributed learning which includes different variation of consensus constraints and the effect of local data sets in each user. We study its solution through variational inequality which gives us a unified view for study this type of problems. We study the existence, uniqueness and distributed algorithm for our proposed setup.}  

\textbf{\textit{ Appendix A.}}
\begin{proof}
	Proof of this Lemma is as follows. Consider the Taylor series expansion of $\widetilde{\mathcal{F}}$ with respect to $\textbf{f}_n$ as $$\widetilde{\mathcal{F}}=[\widetilde{\mathcal{F}}]_{\varsigma_n}+[\sum_{i=1}^\infty (\frac{1}{i!})(\varsigma_n)^i \bigtriangledown_{\textbf{f}_n^i}\widetilde{\mathcal{F}}]_{\varsigma_n}.$$ When $\varsigma_n $ is small, the first term is equal to $0$, and from A2-A3, the second term has a bounded value such that there exists $q^{\text{max}}\leq \infty$ where $\|\mathcal{F}-\widetilde{\mathcal{F}}\| \leq q^{\text{max}}$. Now, we have 
	\begin{equation}\label{strong}
		(\textbf{w}_1-\textbf{w}_2) (\widetilde{\mathcal{F}}(\textbf{w}_1)-\widetilde{\mathcal{F}}(\textbf{w}_2) ) \geq (\textbf{w}_1-\textbf{w}_2) ({\mathcal{F}}(\textbf{w}_1)-{\mathcal{F}}(\textbf{w}_2) )\geq c \|\textbf{w}_1-\textbf{w}_2 \|
	\end{equation}where $c>0$ is a strong monotonicity constant of $\mathcal{F}$, and the last inequality comes from strong monotone assumption of  $\mathcal{F}$.
\end{proof}

\textbf{\textit{ Appendix B.}}
\begin{proof}
	For this theorem, we first should proof the following Lemma. 
	\\	\textbf{Lemma 1}: 1) For the closed, convex uncertainty region in $\Re_n (\textbf{w}_{-n})$, $\widetilde{\textbf{W}}_n$ is a closed, bounded and convex set. 2) $\Phi_n$ is a continuous and differentiable of $\textbf{w}_{n}$ and $\textbf{f}_n(\textbf{w}_{-n}, \textbf{x}_{-n}, \textbf{y}_{-n})$. If $V_n(\textbf{w}_n, \textbf{x}_n, \textbf{y}_n)$ is convex with respect to $\textbf{w}_n$, $\Phi_n$ is a convex function.  
	
	\begin{proof}
		
		This proof has two parts:
		\begin{itemize}
			\item When $\textbf{f}_{n}(\textbf{w}_{-n}, \textbf{x}_{-n}, \textbf{y}_{-n})$ is a linear additive function of other user's wights. Moreover the norm function is a convex function bounded to $\varsigma_n$ (Sec 2.2.2 \cite{boydbook}). Hence $\widetilde{\textbf{W}}_n$ is closed, bounded set.
			\item From A1-A3, $\Phi_n$ is contentious and differentiable. If $V_n(\textbf{w}_n, \textbf{x}_n, \textbf{y}_n)$ is convex, its maximum over the convex and closed set is also convex \citep{boydbook}.
		\end{itemize}
		
	\end{proof}

	From assumptions A1-A4 and Lemma 1, it is straightforward to show Lemma 3.1 and Theorem 3.2 in \citep{existenceofGVI} hold for our setup. Therefore, both $VI(\mathcal{Q}, \mathcal{F})$ and  $VI(\widetilde{\mathcal{Q}}, \widetilde{\mathcal{F}})$ have at least one solution. 
\end{proof}

\textbf{\textit{ Appendix C.}}

\begin{proof}

	From Theorem 2.3.3.(b) in \citep{PangVI}, if $\mathcal{F}+\boldsymbol{\lambda} \mathcal{G}^1+\boldsymbol{\mu}\mathcal{G}_2$ is strongly monotone, the solution of this optimization problem is unique. when $g_n^1$ is linear and $g_n^2$ is convex, we have $$(\textbf{w}-\textbf{w}')(\mathcal{G}^2(\textbf{w})-\mathcal{G}^2(\textbf{w}'))\geq 0$$ and $$(\textbf{w}-\textbf{w}')(\mathcal{G}^1(\textbf{w})-\mathcal{G}^1(\textbf{w}'))\geq 0,$$ and it is straightforward that when $\mathcal{F}$ is strong monotone, $\mathcal{F}^{\text{Fixed}}$ is strong monotone. From Appendix A, and from \eqref{strong} in this supplementary file , when  $\mathcal{F}^{\text{Fixed}}$ is strong monotone, $\widetilde{\mathcal{F}}^{\text{Fixed}}$ is strong monotone. Therefore both $VI(\textbf{W}, \mathcal{F}^{\text{Fixed}})$ and  $VI(\widetilde{\textbf{W}}, \widetilde{\mathcal{F}}^{\text{Fixed}})$ have a unique solution when $\mathcal{F}$ is strong monotone \cite{PangVI}.
\end{proof}

\textit{\textbf{Appendix D} }
\begin{proof}
	For this proof, we first define the P-matrix property of one matrix: 
	
	\textbf{Definition:} For any nonzero vector $\textbf{x}$, we have $x_n(\boldsymbol{\Upsilon}\textbf{x})_n>0$ where $x_n$ is the $n^{\text{th}}$ element of $\textbf{x}$.

	The strong monotone condition of mapping $\mathcal{F}$ can be replaced by a $P$-matrix condition of $\boldsymbol{\Upsilon}$ where \citep{Palomarbook}
	\begin{eqnarray} \label{Upsilon}
		[ \Upsilon]_{nm}= \left\{\begin{array}{c}
			\alpha_{n}^{\text{min}}, \,\,\,\,\,\qquad\text{if} \qquad\, m=n, \\
			-\beta_{nm}^{\text{max}},  \,\qquad  \text{if} \qquad\, m\neq n, \\
		\end{array} \right.
	\end{eqnarray} 
	in which $$\label{alphamin}\alpha_{n}^{\text{min}} \triangleq
	\inf_{\textbf{w} \in \textbf{W}} \alpha_{n}(\textbf{w}),$$$$\beta_{nm}^{\text{max}} \triangleq 
	\sup_{\textbf{w} \in \textbf{W}}\beta_{nm}(\textbf{w}),$$  $$\alpha_{n}(\textbf{w})\triangleq \text{smallest eigenvalue of}  \nabla^{2}_{\textbf{w}_{n}}V_{n}(\textbf{w}),$$ and $$
	\beta_{nm}(\textbf{w}) \triangleq \| \nabla_{\textbf{w}_{n}\textbf{w}_{m}} V_{n}(\textbf{w})\|_2, \quad \forall  n\neq m,$$
	where $\nabla^{2}_{\textbf{w}_{n}}
	V_{n}(\textbf{w})$ and $\nabla_{\textbf{w}_{n}\textbf{w}_{m}}
	V_{n}(\textbf{w})$ are the $K \times K$ Jacobian matrices of
	$\mathcal{F}_n(\textbf{w})$ with respect to $\textbf{w}_n$ and
	$\textbf{w}_m$, respectively, and $\|- \nabla_{\textbf{w}_{n}\textbf{w}_{m}} V_{n}(
	\textbf{w}_n)\|_2$ is the $l_2$-norm of
	$V_{n}(\textbf{w}_{n})$.

	Now for all $n \in \mathcal{N}$, we have $\frac{\partial^2 V_n}{\partial \textbf{w}_n \partial \textbf{f}_n} \ll \frac{\partial^2 V_n}{\partial^2 \textbf{w}_n }$ which indicates that the effect of the weight of user $n$ on its own optimization problem is much stronger than the effect of other users on its own solution, conducting that matrix $\mathcal{J}$ is strictly semipositive matrix where $[\mathcal{J}]_{nn}=\frac{\partial^2 V_n}{\partial^2 \textbf{w}_n }$ and $[\mathcal{J}]_{nm}=\frac{\partial^2 V_n}{\partial \textbf{w}_n \partial \textbf{f}_n } \times \frac{\textbf{f}_n }{\partial \textbf{w}_n }$ and $\frac{\textbf{f}_n }{\partial \textbf{w}_n } \geq 0$.
	Algorithm 1 converges when  $\boldsymbol{\Upsilon}$ is a P-matrix and the third order of derivatives of $V_n$ with respect to $\textbf{w}_n$ are negligible i.e., \textbf{A5} \cite{Palomarbook}.

\end{proof}

\textit{\textbf{Appendix E} }
\begin{proof}
	
	The proof of this theorem include three steps. 
	
	\begin{itemize}
		\item \textbf{Step1:} For any value of fixed $\boldsymbol{\lambda}$, the solution of the mapping (11) for $\textbf{w}$ and $\widetilde{\textbf{w}}$ come from solution of $VI(\textbf{W}, \mathcal{F}^{\text{Fixed}})$ and  $VI(\widetilde{\textbf{W}}, \widetilde{\mathcal{F}}^{\text{Fixed}})$. In Appendix C, we show that when $\mathcal{F}$ is strong monotone, the solution of $VI(\textbf{W}, \mathcal{F}^{\text{Fiexed}})$ and  $VI(\widetilde{\textbf{W}}, \widetilde{\mathcal{F}}^{\text{Fiexed}})$ are unique. Let $(\textbf{w}^*, \boldsymbol{\lambda}^*)$ be a solution of (3) which exists according to Theorem 1. From monotonocinity of $\mathcal{F}$, $\textbf{w}^*$ is unique and $\boldsymbol{\lambda}^*$ is a solution of the NCP in (11). Conversely, if $\boldsymbol{\lambda}^*$ is any solution of the NCP, then $(\textbf{w}^*, \boldsymbol{\lambda}^*)$ should be a solution of (2) \citep{5411806}. The same is true for (2) and  $\widetilde{\textbf{w}}^*$. 
		\item \textbf{Step2:} If NCP in (11) is monotone, there is a convex set of solutions for this equations which interestingly all of them leads to $\textbf{w}^*$. NCP in (11) is monotone, if $\boldsymbol{\phi}$ is co-coercive, i.e., 
		$$\big(\boldsymbol{\lambda}^1-\boldsymbol{\lambda}^2)^T(\boldsymbol{\phi}(\boldsymbol{\lambda}^1)-\boldsymbol{\phi}(\boldsymbol{\lambda}^2))\geq c_{\text{coc}}(\boldsymbol{\phi})\|\boldsymbol{\phi}(\boldsymbol{\lambda}^1)-\boldsymbol{\phi}(\boldsymbol{\lambda}^2)\|_2^2 $$ which conducts us to have Lipschitz continuous property for mapping $\boldsymbol{\phi}$. The same is true for $QVI(\widetilde{\mathcal{Q}}, \widetilde{\mathcal{F}})$.
		\item \textbf{Step3:} To study the co-corrective condition of $\boldsymbol{\phi}$, for any given $\boldsymbol{\gamma}$, let $\textbf{w}^*=\textbf{w}^*(\boldsymbol{\gamma})$ be a unique solution of VI of $\mathcal{F}^{\text{Lagrange1}}(\boldsymbol{\varkappa})$, from $VI(\mathcal{Q}, \mathcal{F}^{\text{Lagrange1}}(\boldsymbol{\varkappa}))$,  we have $$(\textbf{w}^*(\boldsymbol{\gamma}^1)-\textbf{w}^*(\boldsymbol{\gamma}^2)) [\mathcal{F}(\textbf{w}^*(\boldsymbol{\gamma}^2))-\boldsymbol{\gamma}^2)]  \geq 0 $$
		and 
		$$(\textbf{w}^*(\boldsymbol{\gamma}^2)-\textbf{w}^*(\boldsymbol{\gamma}^1)) [\mathcal{F}(\textbf{w}^*(\boldsymbol{\gamma}^1))-\boldsymbol{\gamma}^1)]  \geq 0. $$
		By adding the above two terms and rearranging terms and following the strong monotonicity of $\mathcal{F}$, we have 
		\begin{equation}\label{31}
			(\boldsymbol{\gamma}^1-\boldsymbol{\gamma}^2)(\textbf{w}^*(\boldsymbol{\gamma}^1)-\textbf{w}^*(\boldsymbol{\gamma}^2))\geq c(\mathcal{F})\|\textbf{w}^*(\boldsymbol{\gamma}^1)-\textbf{w}^*(\boldsymbol{\gamma}^2)\|_2^2
		\end{equation}
		where $c(\mathcal{F})$ is strong monotone constant of $\mathcal{F}$. From convexity of $\textbf{g}^2$ and definitions of $\boldsymbol{\phi}$ and $\boldsymbol{\gamma}$, we have $$(\textbf{w}^*(\boldsymbol{\gamma}^1)-\textbf{w}^*(\boldsymbol{\gamma}^2))\geq\frac{1}{\mathcal{G}^2}\boldsymbol{\phi}(\textbf{w}^*(\boldsymbol{\gamma}^1)-\boldsymbol{\phi}(\textbf{w}^*(\boldsymbol{\gamma}^2)$$ and from the definition of $\boldsymbol{\gamma}$, we can show that left term of \eqref{31} in this supplementary file is less than $(\boldsymbol{\lambda}^1-\boldsymbol{\lambda}^2)(\boldsymbol{\phi}(\boldsymbol{\lambda}^1)-\boldsymbol{\phi}(\boldsymbol{\lambda}^2)),$ and therefore, we deduce that $$(\boldsymbol{\lambda}^1-\boldsymbol{\lambda}^2)(\boldsymbol{\phi}(\boldsymbol{\lambda}^1)-\boldsymbol{\phi}(\boldsymbol{\lambda}^2)))\geq \frac{c(\mathcal{F})}{\mathcal{G}^2}\|\boldsymbol{\phi}(\boldsymbol{\lambda}^1)-\boldsymbol{\phi}(\boldsymbol{\lambda}^2)\|_2^2.$$ This inequality proofs the desired co-coercivity property of $\boldsymbol{\phi}$. From this, we can reach to the convex set of the solution of NCP. 
		From Appendix A, the above results can be derived for the $\widetilde{\mathcal{F}}$ with minor modifications. Here, we have $c_{\text{coc}}=\frac{c(\mathcal{F})}{\mathcal{G}^2}$.
	\end{itemize}
	
	Note that this theorem does not show the uniqueness condition of $\boldsymbol{\lambda}$. However, it shows that $\boldsymbol{\lambda}$ belongs to convex set.
\end{proof}

\small
\bibliographystyle{unsrtnat}

\bibliography{myref}





\end{document}